\documentclass[runningheads]{llncs}

\usepackage{amsmath,graphicx}
\usepackage{framed,multirow}

\usepackage{amssymb}
\usepackage{latexsym}

\usepackage{subcaption}
\usepackage{caption}
\usepackage{xcolor}
\usepackage{graphicx}

\usepackage{makecell}
\usepackage{array}
\usepackage{pgfplots}

\usepackage{multirow}
\usepackage{todonotes}
\newcommand\MyBox[2]{
  \fbox{\lower0.75cm
    \vbox to 1.4cm{\vfil
      \hbox to 1.4cm{\hfil\parbox{1.2cm}{#1\\#2}}
      \vfil}%
  }%
}
\newcommand\MyBoxx[2]{
  \mbox{\lower0.5cm
    \vbox to 1cm{\vfil
      \hbox to 1cm{\hfil\parbox{1cm}{#1\\#2}\hfil}
      \vfil}%
  }%
}

\usepackage{url}
\usepackage{xcolor}
\definecolor{newcolor}{rgb}{.8,.349,.1}

\begin{document}
\title{An Ambient Intelligence-based Approach For Longitudinal Monitoring of Verbal and Vocal Depression Symptoms }
%
%
\author{Alice Othmani\inst{1,2}\orcidID{ 0000-0002-3442-0578}$\spadesuit$ \and
Muhammad Muzammel\inst{1}}

\authorrunning{A. Othmani et al.}
%
\institute{Université Paris-Est Créteil (UPEC), LISSI, Vitry sur Seine, 94400, France \\
\email{\{alice.othmani,muhammad.muzammel\}@u-pec.fr} \\
$\spadesuit$ Corresponding author: \email{alice.othmani@u-pec.fr}}

\maketitle              
\begin{abstract}
 Automatic speech recognition (ASR) technology can aid in the detection, monitoring, and assessment of depressive symptoms in individuals. ASR systems have been used as a tool to analyze speech patterns and characteristics that are indicative of depression. Depression affects not only a person's mood but also their speech patterns. Individuals with depression may exhibit changes in speech, such as slower speech rate, longer pauses, reduced pitch variability, and decreased overall speech fluency. 
 Despite the growing use of machine learning in diagnosing depression, there is a lack of studies addressing the issue of relapse. Furthermore, previous research on relapse prediction has primarily focused on clinical variables and has not taken into account other factors such as verbal and non-verbal cues.
Another major challenge in depression relapse research is the scarcity of publicly available datasets. To overcome these issues, we propose a one-shot learning framework for detecting depression relapse from speech. We define depression relapse as the similarity between the speech audio and textual  encoding of a subject and that of a depressed individual. To detect depression relapse based on this definition, we employ a Siamese neural network that models the similarity between of two instances. Our proposed approach shows promising results and represents a new advancement in the field of automatic depression relapse detection and mental disorders monitoring.

\end{abstract}
\section{Introduction}\label{sec1}

Major Depressive Disorder (MDD) is a mood disorder that has detrimental effects on an individual's cognition, emotions, and daily functioning. It is primarily characterized by persistent feelings of sadness, anger, and loss of interest in activities. MDD is among the most prevalent mental disorder, impacting over 300 million people globally  \cite{Marcus2012}. Furthermore, recent studies have indicated a significant rise in mental health issues, including anxiety, stress, and depression, during the COVID-19 pandemic \cite{shah2021prevalence}.

Depression relapse refers to the re-occurrence of depressive symptoms after a period of partial or complete remission. It means that a person who previously experienced an episode of depression and showed improvement or recovery from their symptoms subsequently experiences a return or worsening of those symptoms \cite{monroe2011}. Relapse can happen during or after treatment for depression, and it is often characterized by a recurrence of the emotional, cognitive, and behavioral manifestations associated with depression.  
 Relapse and recurrence rates are high in MDD patients with a percentage of $60\%$ after 5 years, $67\%$ after 10, and $85\%$ after 15 \cite{hardeveld2010}. Thus,  there  is  a  pressing  need for automatic monitoring systems  which  can be employed to detect depression relapse or recurrence at an early stage, thereby facilitating timely intervention. 

Despite the increasing utilization of machine learning (ML) for analyzing Major Depressive Disorder (MDD) and other mental health issues, and its potential to enhance decision-making for mental health practitioners, \cite{muzammel2021end,dwyer2018,gao2018,su2020}, there is a noticeable lack of studies utilizing ML to address the issue of depression relapse. Additionally, previous research on relapse prediction primarily relies on clinical variables such as age, gender, medication types, number of episodes, symptom severity, cognitive markers, and medical image data \cite{borges2018,chanda2020,sato2015,ruhe2019}. However, these approaches have overlooked other important attributes, such as the analysis of speech patterns and facial expressions. 

On the other hand, a big limitation to depression research studies is the lack of public datasets. Moreover, due to privacy, safety, expense, and ethical concerns \cite{thelisson2018general}, acquiring examples to train a model for depression relapse is a difficult task. This motivates research for approaches that take into account the data scarcity problem in this area. 
Few-shot learning is a subfield of machine learning that deals with the challenge of learning new concepts or tasks with limited labeled training data. In traditional machine learning approaches, a large amount of labeled data is typically required to train models effectively. However, in few-shot learning, the goal is to develop algorithms that can generalize and learn from only a few examples or instances of a particular class or task.

 In this paper, we propose a robust approach that deals with depression relapse data scarcity. The proposed approach is based on one-shot learning. We define depression relapse as the closeness of the speech encoding of a subject to that of a depressed subject.
By modeling the similarity between two instances, Siamese neural network is chosen as a suitable candidate for depression relapse detection following the proposed definition.
The proposed approach investigates the predictive power of audio and textual encodings of the speech for depression relapse prediction using Siamese neural network architecture.  Three Siamese networks built using (1) MFCC audio features, (2) VGGish audio features, and (3) audio-textual fusion features are compared and investigated for the relapse identification task. To our knowledge, this work is the first to tackle the prediction of depression relapse based on verbal and non-verbal cues in a speech, and to employ one-shot learning for this task.

The paper is structured as follows. In the following section, a literature review is conducted to identify major previous works done to predict depression relapse and recurrence. An overview of n-shot learning and Siamese networks is also included (section \ref{sec2}). 
In section \ref{sec3}, the different steps of our methodology are  described. Finally, the obtained results and discussion of the performed tests are presented in section \ref{sec4}.

\section{Related work}\label{sec2}
Relapse and recurrence can be attributed to various factors. It is suggested that recurrence is ascribed to a genetic vulnerability \cite{burcusa2007}. Likewise, some findings suggest that inflammation can issue a possible mechanism for recurrent depression \cite{liu2019}. Another factor can be neuropsychological functioning 
\cite{ruhe2019}. 
Factors triggering depression relapse may vary among individuals. Moreover,  in many cases it is difficult to underpin the trigger factors leading to a depression relapse episode. This motivates the development of automatic monitoring systems capable of detecting relapse.

Researcher for predicting depression relapse often relies on n clinical variables such as medication type, episode number, symptom severity, cognitive markers, and occasionally medical imaging data. For example, Borges et al. (2018) \cite{borges2018} have explored the use of machine learning algorithms in predicting relapse in bipolar patients based on their clinical data, with Random Forests showing promising results (68\% for the Relapse Group and 74\% for the No Relapse Group). Another study by Chanda et al. (2020) \cite{chanda2020} aimed to develop a recurrence classification platform based on gender, age, medication, and treatment time, where K-Nearest Neighbor outperformed SVM and RF with an 83\% accuracy. Emotional biases were investigated  investigated by Ruhe et al. (2019) \cite{ruhe2019}, as potential biomarkers for depression relapse, resulting in a linear SVM model with a 75\% accuracy. Moreover, a multimodal approach proposed by Cearns et al. (2019) \cite{cearns2019} incorporating various modalities, including clinical, blood-biomarker, genetic, bio-electrical impedance, electrocardiography, and structural imaging, achieved an accuracy of 65.72\% using SVM. Further, a holistic approach involving the analysis of depression scores trajectories and predicting individual treatment outcomes was proposed using smoothing splines, K-means clustering, and collaborative modeling \cite{lin2016}. 

Just recently, three research studies have emerged that propose utilizing video data for the prediction of depression relapse \cite{othmani2022model,muzammel2021identification,othmani2022multimodal}.
 In \cite{othmani2022model} along with CNN, a Model of Normality (MoN) was utilize to detect depression and relapse. While, the other approach \cite{muzammel2021identification} proposed a preliminary study based one-shot learning due to capacity to learn instantly. Both of these state of the art modalities relies on audio and visual features for  monitoring depression relapse. In \cite{othmani2022multimodal}, a deep learning-based approach is proposed for depression recognition and depression relapse prediction using videos of clinical interviews. Their approach is based on a correlation-based anomaly detection framework and a measure of similarity to depression where depression relapse is detected when the deep audiovisual patterns of a depression-free subject become close to the deep audiovisual patterns of depressed subjects. Thus, the correlation between the audiovisual encoding of a test subject and a deep audiovisual representation of depression is computed and used for monitoring depressed subjects and for predicting relapse after depression.

\section{Methodology}\label{sec3}

Depression relapse is defined in this work as the similarity (dissimilarity) between the audio-textual speech encoding of a subject and the speech encoding of a diagnosed depressed (non-depressed) subject. One-shot learning based Siamese network is chosen in this work  for modeling depression relapse, as it models the similarity (dissimilarity) between two samples.
 The proposed framework is composed of four stages: (1) pre-processing audio data augmentation (section~\ref{subsec3.1}), (3) audio-textual features extraction (section~\ref{subsec3.3}), and (4) one-shot learning-based depression relapse detection (section ~\ref{subsec3.4}). Each of these steps are detailed in the following. 
\begin{figure*}[]
    \centering
        \includegraphics[width=\linewidth]{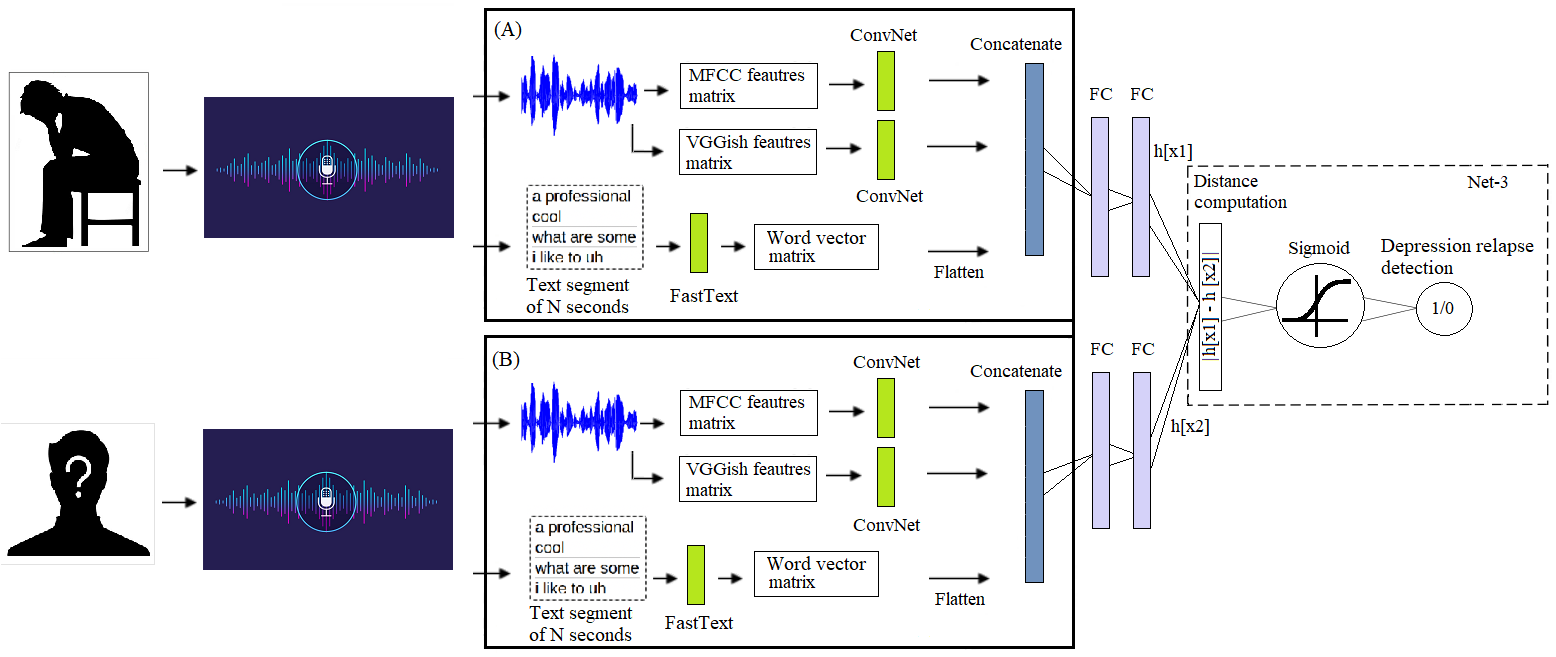}
\caption{Proposed multimodal-based Siamese networks for depression relapse detection.}
\label{fig:Net}
\end{figure*}

\subsection{Pre-processing \& Data Augmentation}
\label{subsec3.1}

\subsubsection{Audio Preprocessing} 
In the pre-processing step, unvoiced segments are first filtered and the speech of the subject is extracted from the audio signal. The speech signal is then divided into $n=7.6$ seconds speech segments.
To increase the number of data samples and the  system's robustness to noise, data augmentation is performed. The signals are perturbed using two audio augmentation techniques \cite{rejaibi2019mfcc,muzammel2020,Othmani2021}:

1)\textit{ Noise Injection}: the audio signal is perturbed through the injection of random noise. Let $x$ be an audio signal and $\alpha$ a noise factor, the noise perturbed signal $x_N$ is given by:
$x_N=x-{{\alpha}\times{rand(x)}}$, with $\alpha$ = $0.01$, $0.02$ and $0.03$.

2) \textit{Pitch Augmentation}:  pitch is lowered by factors of $0.5$, $2$, and $2.5$ (in semitones).

\subsubsection{Text Preprocessing}
The clinical interviews are recorded and accompanied by transcriptions of the conversations between the participant and the interviewer. Unlike the audio data, we analyze both the transcriptions of the interviewer and the participant. This is because the verbal reactions of the interviewer following the participant's responses can contain valuable information about the participant's emotions. For example, when the participant responds negatively, the interviewer may express phrases like "that sucks" or "I'm sorry to hear that," which provide insight into the participant's depressive state.
Our decision to focus solely on the audio patterns of the participant is based on the fact that the interviewer's audio patterns do not exhibit signs of depression. However, the words used by the interviewer can indicate sympathy when the patient is depressed.

\subsection{Audio-Textual Features Extraction}\label{subsec3.3}
Following data augmentation, audio-textual features are extracted. These constitute MFCC and Vggish audio features, and textual word2Vec features.

\subsubsection{MFCC Features Extraction from Audio Signal}\label{subsubsec3.3.1}

MFCC features represent the audio cepstrum in the non-linear Mel scale. Such representation is said to approximate the human auditory system. 
Below is  the detailed description of the MFCC feature extraction  steps.

\textit{Windowing}: The signal is  split into  $60$ milliseconds frames. A Hamming window is then applied on each frame to taper the signal towards the frame boundaries.  Given a signal $s[n]$ of length $n$  and a hamming window $w[n]$, the sliced frame is given by:

\begin{equation}
x[n] = w[n]s[n] \quad with \quad w[n]=\alpha -\beta\cos\left(2\pi\frac{n} {N-1}\right) 
\end{equation}


where  $\alpha=0.54$, $\beta=0.46$, $N$ is the window length such that $0 \leq n \leq N$.\\

\textit{DFT spectrum}: Discrete Fourier Transform (DFT) is then performed on each windowed frame  to get the magnitude spectrum:

\begin{equation}
X[k]=\sum_{n=0}^{N-1}{x[n]}e^{-j\frac{2\pi}{N}kn}; \quad 0 \leq k \leq N-1
\end{equation}

\textit{Mel spectrum}: Triangular Mel-scale filter banks are then multiplied by the magnitude spectrum to compute the Mel spectrum.   

\begin{equation}
Y_t[m]=\sum_{k=0}^{N-1}{W_m[k]|X_t[k]|^2}; \quad 0 \leq m \leq M-1
\end{equation}
where $W_m$ represents the $m^{th}$ triangular Mel-scale filter bank,  $M$ is total number of filters, and $k$ corresponds to the DFT bin number.\\

\textit{Discrete cosine transform (DCT)}: The Mel spectrum is then represented on a log scale and DCT is  applied generating a set of cepstral coefficients. MFCC is thus calculated as:

\begin{equation}
c[n]=\sum_{m=0}^{M-1}{log_{10}\left(Y_t[m]\right)cos\left(n(m-0.5)\frac{\pi}{M}\right)}
\end{equation}

For this work, 60 dimensional MFCC features are extracted leading to a matrix of $378\times60$ for each $7.6$ seconds signal. \\

\subsubsection{Vggish Features Extraction from Audio signal}\label{subsubsec3.3.2}

In this work, VGGish features \cite{hershey2017} are extracted from the audio segments. VGGish converts audio features into semantically significant compact high-level 128 dimensional embedding. This embedding can be then fed  to a shallow or deep classification model.  

To compute the VGGish features, first a one-sided Short-Time Fourier Transform (STFT) is applied to the audio clip using a 25 ms periodic Hann window with 10 ms hop, and 512-point DFT. 
The complex spectral values are then converted to magnitude and  phase information is discarded.
The one-sided magnitude spectrum are passed to a 64-band mel-spaced filter bank, and the magnitudes in each band are then summed. The obtained mel spectrograms are then converted into a log scale and buffered into overlapped segments consisting of 96 spectrums each. These mel spectrograms are passed through VGGish network to obtain  a 14x128 matrix \cite{gemmeke2017,hershey2017}.  

The VGGish feature matrix contained zero values. To resolve this issue, the VGGish model was released with a precomputed principal component analysis (PCA) matrix \cite{gemmeke2017,hershey2017}. First, we subtract the precomputed 1x128 PCA mean vector from the 14x128 feature matrix, and then premultiply the result by the precomputed 128x128 PCA matrix.

\subsubsection{Textual Features Extraction}\label{subsubsec3.3.3}
For each audio segment ($7.6sec$) of the clinical interviews, the words spoken by both the participant and the interviewer are converted into sequences of vectors using word embedding as a textual features.
Word embedding is a technique that transforms words into dimensional vectors, where words with similar meanings or words that frequently occur together in context are represented by vectors that are close to each other in the dimensional space. In particular, each frame transcript is represented by a matrix $E=(e_1, \dots, e_k, \dots e_{nw})$ where $e_k$ is the word vector corresponding to the $k^{th}$ word and $nw$ is the number of words in the frame transcript.  

For word embedding, we utilize the fastText pretrained network \cite{mikolov2018advances}, which was trained on Common Crawl \footnote{https://commoncrawl.org/2017/06} and incorporates sub-word information. This network produces word vectors of size $300$. In cases where certain words are not present in the pretrained model, we replace them with their synonyms. Additionally, the resulting word vector matrix for each transcript frame is resized to $60\times9$, where $60$ represents the size of MFCC coefficients and $9$ represents the minimum number of words present in a single frame.

\subsection{One-Shot Learning Framework for Depression Relapse Detection}\label{subsec3.4}
In this work, 1D convolutional Siamese networks are proposed for depression relapse detection based on audio-textual cues. The proposed architectures are summarized in Fig.~\ref{fig:Net}. 

\subsubsection{One-shot learning }
One-shot learning refers to the classification task where the number of available instances per class is limited. In some cases, a class may have only a single example during the model training phase. The objective of one-shot learning is to train models that can determine the similarity between a pair of inputs and identify whether they belong to the same class or not. 

Siamese neural networks, also known as twin neural networks, are a type of one-shot learning model that aim to learn a distance function between pairs of input vectors. In the architecture of a Siamese neural network, two identical sub-networks with shared weights are employed. Each sub-network takes in a distinct pair of inputs and produces output representations. The network then computes a distance metric, such as Euclidean distance or cosine similarity, between the outputs of the sub-networks. This distance metric reflects the similarity between the two input vectors \cite{Chicco2020}. The training can be done on different pairs of inputs from all possible classes. To obtain a prediction for an instance, a comparison between the instance and reference instances of all classes should be performed. The final prediction is then obtained based on the different similarity scores.

For training a Siamese network, dataset pre-processing is required. A pairing procedure is performed where pairs of data points are created: similar samples pair, and dissimilar samples pair. Similar samples are then assigned positive labels, while dissimilar samples are assigned negative labels. The pairs are then fed to the Siamese architecture.  The pairing procedure employed in this paper is detailed in section \ref{subsec:dataset}.

\subsubsection{Proposed Siamese Architectures}
We propose two Siamese networks built using (1) MFCC audio features, (2) VGGish audio features, and (3) audio-textual fusion features. 

\textbf{Audio based Siamese Network:} The proposed MFCC and VGGish Siamese models have similar architectures. The architecture  consists of two blocks of convolutional, Relu, dropout, and flatten layers. 
Two fully connected layers are used to compute encodings and an euclidean distance is measured between the two encodings. In other words, both Net-1 (A and B) from Fig. \ref{fig:Net} are connected with Net-3 through the fully connected layers.
A last layer of size two is added with a sigmoid activation function for binary prediction of relapse.

\textbf{Audio-Textual based Siamese Network:}
In Fig. \ref{fig:Net}, the 60 MFCC features are fed to a Convolutional Neural Network composed of two blocks of 1D convolutional and Relu layers, followed by dropout and flatten layers. Similarly 14 VGGish features are fed to a similar Convolutional Neural Network consisting of two blocks of 1D convolutional and Relu layers, followed by dropout and flatten layers. 
These high level MFCC and VGGish features are then concatenated with the textual features into a features vector of size $540$ and fed to two fully connected layers to obtain the audio-textual encoding. Afterwords, an euclidean distance is computed between the two encodings: one of non-diagnosed subject and one of a reference depressed subject. A last layer of size two is also added for binary prediction of relapse. 

\section{Results and Discussion}\label{sec4}

\begin{figure*}[]

\resizebox{\textwidth}{!}{
\subfloat[MFCC Siamese]{
\begin{tabular}{c >{\bfseries}r @{\hspace{0.7em}}c @{\hspace{0.4em}}c @{\hspace{0.7em}}l}
  \multirow{10}{*}{\rotatebox{90}{\parbox{1.1cm}{\bfseries\centering Predicted}}} & & \multicolumn{2}{c}{\bfseries Actual} &\\
  & & \bfseries NS & \bfseries S & \bfseries total \\
  & NS & \MyBox{356}{(33.46)} & \MyBox{187}{(17.58)} & \MyBoxx{65.56}{34.44}  \\[2.4em]
  & S & \MyBox{176}{(16.54)} & \MyBox{345}{(32.42)} & \MyBoxx{66.22}{33.78}  \\
  & total & \MyBoxx{66.92}{33.08} &  \MyBoxx{64.85}{35.15} &
\end{tabular}}
\subfloat[VGGish Siamese]{
\begin{tabular}{c >{\bfseries}r @{\hspace{0.7em}}c @{\hspace{0.4em}}c @{\hspace{0.7em}}l}
  \multirow{10}{*}{} & 
    & \multicolumn{2}{c}{\bfseries Actual} & \\
  & & \bfseries NS & \bfseries S & \bfseries total \\
  & NS & \MyBox{358}{(33.65)} & \MyBox{183}{(17.20)} & \MyBoxx{66.17}{33.83}  \\[2.4em]
  & S & \MyBox{174}{(16.35)} & \MyBox{349}{(32.80)} & \MyBoxx{66.73}{33.27}  \\
  &  & \MyBoxx{67.29}{32.71} &  \MyBoxx{65.60}{34.40} &
\end{tabular}}
\subfloat[Audio-Textual Siamese]{
\begin{tabular}{c >{\bfseries}r @{\hspace{0.7em}}c @{\hspace{0.4em}}c @{\hspace{0.7em}}l}
  \multirow{10}{*}{} & 
    & \multicolumn{2}{c}{\bfseries Actual} & \\
  & & \bfseries NS & \bfseries S & \bfseries total \\
  & NS & \MyBox{404}{(37.97)} & \MyBox{158}{(14.85)} & \MyBoxx{71.89}{28.11}  \\[2.4em]
  & S & \MyBox{128}{(12.03)} & \MyBox{374}{(35.15)} & \MyBoxx{74.50}{25.50}  \\
  &  & \MyBoxx{75.94}{24.06} &  \MyBoxx{70.30}{29.70} & 
\end{tabular}}}
\caption{Confusion matrices of Siamese networks for depressive state similarity detection. NS: Non-Similar, S: Similar.}
\label{fig:confusion}
\end{figure*}
\subsection{Dataset}
\label{subsec:dataset}
We use the Distress Analysis Interview Corpus Wizard-of-Oz dataset (DAIC- WOZ) (\cite{gratch2014distress}). This dataset was introduced in AVEC 2017 challenge \cite{ringeval2017}. $189$ subjects took part in the data collection, where they were interviewed by a virtual agent manipulated in a wizard-of-oz setting. The subject's speech and interviews transcripts were saved and publicly made available for research.  The average length of the audio samples is $15$ minutes, sampled at $16 kHz$.
The dataset includes depression scores in two formats: binary and severity level. Binary depression and severity labels were collected  via a self-report Patient Health Questionnaire (PHQ) \cite{williams2005}.
Only $182$ subjects were used in the evaluation. We use a percentage split strategy where the  dataset is randomly divided into $80\%$ training, $10\%$ validation, and $10\%$ testing. Out of the $182$ subjects, the training dataset included $146$ subjects, while the validation and test sets included $18$ subjects each.

\textbf{Pairing} 
Features pairs are then generated to train, test and validate the one-shot learning model. Samples from the train set are randomly paired with each others to form the training pairs. No sample was paired with itself. 
To generate the test and validation pairs, samples from these sets are paired with the samples from the training set.     
The binary depression score provided in DAIC-WOZ consists of two classes: depressed and non-depressed. Binary depression relapse is defined to be the paired feature groups of the couple (Non-Depressed, Depressed). Also, in both classes (i.e., Depressed, Non-Depressed) 50\% features matrix groups are paired with the same class, while the remaining $50\%$ features matrix groups are paired with the other class. Also pairs are created for PhQ-score provided in DAIC-WOZ dataset. For all 24 classes $50\%$ features matrix groups are paired with the same class, while the remaining $50\%$ features matrix groups are paired with the other classes.

\subsection{Network Implementation Details}
A ReLU activation function with $64$  filters is used for the convolutional layers in the presented CNN architectures. The stride and filter sizes are set to $1$ and $3$, respectively. The dropout fraction value is set to $0.01\%$. The dense layers' sizes are set to $1024$ and hyperbolic tangent (tanh) is used as an activation function. To predict the similarity between two feature sets for depression relapse detection based on PHQ-binary, the output layer is a dense layer with a sigmoid activation function and a size of $2$. To predict the similarity based on PHQ-score pairs, the output dense layer has a size of $25$. 
To train the models, an initial learning rate of $10^{-5}$ and a decay of $10^{-6}$ is used. The batch size and epochs are set to $100$ and $300$, respectively. For both models, Root Mean Square Error is used as a loss function and trained with RMSProp optimizer. An early stopping is used if the loss function stops decreasing after 10 epochs.

\begin{table}
\caption{Performance of proposed Siamese networks for depressive state similarity in terms of accuracy, Root Mean Square Error (RMSE) and CC}\label{tab1}
\centering
\begin{tabular}{|l|l|l|l|}
\hline
\textbf{ Network } & \textbf{Acc. (\%)} &  \textbf{RMSE} &  \textbf{CC} \\ 
\hline \hline
MFCC Siamese  & 65.88 & 0.4841 & 0.3177 \\
\hline
VGGish Siamese  & 66.45 & 0.4792 & 0.3290 \\
\hline
Audio-Textual Siamese &  \textbf{73.12} & \textbf{0.4585} & \textbf{0.4631} \\
\hline
\end{tabular}

\end{table} 

\subsection{Performance Analysis of Siamese Networks}

The proposed Siamese networks are  evaluated using accuracy, Root Mean Square Error (RMSE), and Pearson Correlation Coefficient (CC) metrics. The proposed framework reaches an accuracy of $65.88\%$ and $66.45\%$ when using only the MFCC features and VGGish features, respectively. The fusion of MFCC, VGGish and textual features notably increases the performance where an accuracy of $73.12\%$ is obtained. A minor decrease in RMSE and a minor increase in CC is noted for the VGGish model compared to the MFCC based one. Further, the CC value of MFCC and VGGish based Siamese networks are $0.3177$ and $0.3290$, which increases to $0.4631$ with the fusion of audio and textual features.

Fig. \ref{fig:confusion} shows the confusion matrices of MFCC, VGGish and Audio-textual based Siamese networks. From the figure, one can notice that the fusion of textual features with audio features network adequately improves the performance of one shot learning Siamese network. For non-similar 
feature pairs classification, the audio-textual based fusion network achieved $9.02\%$ and $8.65\%$ better results compared to MFCC and VGGish based networks. Also for non-similar feature pairs, a notable decrement in false positives has been reported for audio-textual based fusion network.  

For similar feature pairs (i.e., when both features sets in pair belong to the same class) classification the Audio-Textual based fusion network obtained $5.45\%$ and $4.70\%$ better results compared to MFCC and VGGish based networks. Also a considerable decrement in false positives has been reported for Audio-Textual based fusion network compared to MFCC and VGGish based Siamese networks. Furthermore, We also investigate the pair matching for PhQ-Score using multi-class Audio-Textual based Siamese network and we obtained an RMSE value of 4.025 (Normalized RMSE of 0.161).

\section{Conclusion and Future Work}\label{sec5}

In this work, an ASR framework is proposed for depression relapse detection, modeling the similarity of audio and textual speech encoding between a new subject and a diagnosed depressed subject using one-shot learning. The proposed model gave reliable results using the speech's audio and textual cues. The fusion of audio and textual features enhanced the one-shot learning model performance, which made it reliable for detecting depression relapse. Further, the proposed ASR system could help depression patients to monitor their recovery. Lastly, in future work we plan to consider also visual cues in the proposed framework.

\section{ACKNOWLEDGMENTS}
\label{sec:ack}
This work is funded under grant number IF040-2021 (MATCH2021: Malaysia France Bilateral Research Grant).

\bibliographystyle{splncs04}
\bibliography{refs}

\end{document}